\documentstyle[preprint,aps]{revtex}
\tighten
\begin{document}
\preprint{UTHEP-301, SNUTP 95-059, May 1995}
\draft
\flushbottom

\title{\vspace*{1.5in}
QCD Sum Rules, Scattering Length \\
 and the Vector Mesons  in Nuclear Medium}

\author{Tetsuo Hatsuda$^{1,2}$, Su Houng Lee$^{2,3}$, and
 Hiroyuki Shiomi$^1$}

\address{$^1$ Institute of Physics, University of Tsukuba, Tsukuba,
 Ibaraki 305, Japan \\
 $^2$ Institute for Nuclear Theory, NK-12, Univ. of Washington,
 Seattle, WA 98195, USA \\
$^3$ Department of Physics, Yonsei University, 
  Seoul 120-749, Korea }

\maketitle



\vspace{1.5cm}

\begin{abstract}

 Critical examination is made on the relation between 
 the mass shift of 
 vector mesons in nuclear medium and the vector-meson $-$ nucleon
 scattering length.  We give detailed 
 comparison between the QCD sum rule approach by two of the
 present authors (Phys. Rev. {\bf C46} (1992) R34) 
  and the scattering-length
 approach by Koike (Phys. Rev. {\bf C51} (1995) 1488).
  It is  shown that the latter approach is mortally flawed both
 technically and conceptually.

\end{abstract}

\vspace{0.5cm}


\newpage

\section{Introduction}

The question of changes in the vector meson properties in nuclear
 medium is of 
interest in relation to the physics of nuclear matter\cite{Gerry} 
and relativistic 
heavy ion collision \cite{RHIC}.   
In particular, if there are spectral changes of vector mesons 
 in medium, it could be observed directly 
  through the lepton-pair spectrum in $\gamma-A, p-A$ and $A-A$
 reactions \cite{review}.   The problem has been  
studied in effective hadronic models and QCD sum rules (QSR) 
generalized to finite baryon density.  
In QSR, it was shown that 
  the vector meson mass  drops to 
about  10$-$20 \% of its vacuum value in nuclear matter 
density\cite{HL92}.   
This is consistent with the idea of the 
 Brown-Rho scaling \cite{BR91} and also  with Walecka model 
calculations including the  
 vacuum polarization \cite{Florida}.    

Recently, Koike\cite{Koike93} 
claimed that the previous QSR calculation for vector meson in
 nuclear medium\cite{HL92} is not correct 
and that the vector meson mass should increase slightly. 
 Since both ref.\cite{HL92} and \cite{Koike93} on the vector meson 
masses are based on the Fermi 
gas approximation for nuclear matter, it is desirable to clarify the 
origin of the difference.
In this work, we will first give a correct account of the Fermi-gas 
approximation of the QSR in medium 
and show  dangers to relate mass 
shifts  with the scattering length.  Secondly, we will show that the 
 approach in \cite{Koike93} is  {\it technically} wrong so that 
in a correct application of QSR in medium\cite{HL92} the 
vector meson mass does decrease in nuclear medium.

This paper is organized as follows.  In section II, we will
  summarize the results of  ref.\cite{HL92} and 
 make  new remarks which are not given in \cite{HL92} but are
 relevant  for the later discussions.  In section III, 
 the essential points of the Koike's claim are
  summarized.
 Section VI is devoted to critical investigation of \cite{Koike93}.
 We will show that his method is 
 mortally flawed. 
 In section V,  further
 elaboration of ref.\cite{HL92} is discussed.

\vspace{1cm}

\section{QSR in nuclear medium}

Let's start with  the retarded current correlation function in 
nuclear matter,
\begin{eqnarray}
\label{correlator}
\Pi^R_{\mu \nu} (\omega , {\bf q} ;n_{_B})
=i \int d^4x e^{iqx}  \langle {\rm R} J_{\mu}(x) J_{\nu}(0)
 \rangle_n \ \ ,
\end{eqnarray}
where  $q^\mu \equiv (\omega , {\bf q})$ and 
 ${\rm R} J_{\mu}(x) J_{\nu}(0) \equiv 
  \theta(x^0) [J_{\mu}(x), J_{\nu}(0)]$ with the source
currents $J_\mu$ defined as $J_\mu 
={1\over2}(\bar{u}\gamma_\mu u \mp \bar{d}\gamma_\mu d)$ 
($- (+)$ is for the $\rho^0 (\omega)$-meson).
  $n_{_B}$ denotes the baryon number density and 
 $\langle \cdot \rangle_n$ is the expectation
 value in the ground state of nuclear matter.

 Although there are two independent
invariants in medium (transverse and longitudinal polarization),
 they coincide  in the limit  ${\bf q} \rightarrow 0$
 and reduce to
 $\Pi^R_{\mu \mu}/(-3\omega^2) \equiv \Pi^R$.
 $\Pi^R $ satisfies the following dispersion
relation,
\begin{eqnarray}
\label{dispersion}
{\rm Re} \Pi^R (\omega^2) =
 {1 \over \pi} {\rm P} \int_0^{\infty} du^2
{ {\rm Im} \Pi^R(u) \over u^2-\omega^2} + ({\rm subtraction}). 
\end{eqnarray}
In QSR, the spectral density ${\rm Im} \Pi^R$
 is  modeled with several 
phenomenological parameters, while  ${\rm Re} \Pi^R$
  is calculated using the 
operator product expansion (OPE).
 The phenomenological parameters are then  extracted  
by matching the left and right hand side of (\ref{dispersion})
 in the asymptotic region $\omega^2 \rightarrow - \infty $.

 Although the nuclear matter ground state has complicated structure,
 the Fermi-gas approximation taking into account the Pauli principle
 among the nucleons is a good starting point.
 In this approximation, $\Pi^R_{\mu \nu}$ reads
\begin{eqnarray}
\label{fermicor}
\Pi^R_{\mu \nu} (\omega , {\bf q} ;n_{_B})
= \Pi^R_{\mu \nu} (\omega , {\bf q} ;0)
+ \gamma  \int^{p_f} {d^3 p \over (2\pi)^3 2E_N} 
 T^R_{\mu \nu} (\omega, {\bf q} \mid {\bf p}) ,
\end{eqnarray}
with
\begin{eqnarray}
\label{Tmatrix}
 T^R_{\mu \nu} (\omega, {\bf q}\mid {\bf p}) 
 = i \int d^4x e^{iqx}\langle N({\bf p})|
 {\rm R} J_{\mu}(x) J_{\nu}(0) | N({\bf p})
  \rangle .
\end{eqnarray}
Here $\gamma$ is a degeneracy factor ($\gamma$=4 in nuclear matter),
 $|N({\bf p}) \rangle$ is the nucleon state with four momentum
 $p^{\mu}=(E_N,{\bf p})$ in the covariant normalization
 $\langle N({\bf p}) |N({\bf p}') \rangle
=(2\pi)^3 2E_N \delta^3( {\bf p}-{\bf p'})$. 
 The spin average for the nucleon state is assumed in
 (\ref{Tmatrix}).
 $ T^R_{\mu \nu} (\omega, {\bf q} \mid {\bf p})$ is nothing but the
 spin-averaged forward scattering amplitude of the external current
 $J_{\mu}$  and the nucleon.  In nuclear matter, ${\bf p}$ is
 integrated  out in the
 range $0 < \mid {\bf p}\mid  < p_f$ ($p_f$ is the fermi momentum).

The OPE for ${\rm Re} \Pi^R(\omega^2)$, which is the same with 
 the OPE for the causal (Feynman) correlation  $\Pi^F(\omega^2)$,
 has a general form at $\omega^2 \equiv - Q^2 \rightarrow - \infty$,
\begin{eqnarray}
\label{ope}
 {\rm Re} \Pi^R(\omega^2 \rightarrow - \infty)
 = \sum_i \frac{1}{Q^{2i}} a_i(Q^2;\mu^2)
 \langle {\cal O}_i (\mu^2) \rangle_n \ \ \ ,
\end{eqnarray}
where $\mu$ is the renormalization point of the local  operators,
 which separates the  hard scale $|\omega|$ and soft scales such 
as  $\Lambda_{QCD}$ and $p_f$.
 The Wilson coefficients do not depend on  the medium effect
 and only  the expectation 
 values  $\langle {\cal O}_i (\mu^2) \rangle_n$  
 have the  $p_f$
 dependence.

The local operators ${\cal O}_i(\mu^2)$
 in the vector meson sum rule are essentially 
 the same with those 
 in the lepton-nucleon deep inelastic scattering (DIS) and
 can be characterized by their canonical dimension ($d$) and the 
twist ($\tau$=dimension-spin).   They are given in \cite{HKL93}
  up to dimension 6 operators and we will not recapitulate them here.
 Since we are taking ${\bf q} \rightarrow 0$,
 eq. (\ref{ope}) is  an asymptotic series in $1/{\omega}^2$
 or equivalently  an expansion with respect to  $d$.

Now let us examine the density dependence of 
$ \langle {\cal O}_i (\mu^2) \rangle_n $.
 In the Fermi-gas approximation, it reads
\begin{eqnarray}
\label{Fermiop}
\langle {\cal O} \rangle_n= \langle {\cal O} \rangle_0 +
 \gamma  \int^{p_f} {d^3 p \over (2 \pi)^3 2E_N} \langle N({\bf p})| 
 {\cal O} | N({\bf p}) \rangle.
\end{eqnarray}

For the scalar matrix elements such as $\langle N |\bar{q}q| N
 \rangle$ and $\langle N |G^2| N \rangle$,
 one can utilize the information of 
 the $\pi-N$ sigma term $\Sigma_{\pi N}$ 
 and  the QCD trace anomaly \cite{HK92}. The results are
\begin{eqnarray}
\langle \bar{u}u \rangle_n= \langle \bar{u} u \rangle_0  +
 { \Sigma_{\pi N} \over 2 \hat{m} } \cdot n_{_B}  \cdot B_1(x), 
 ~~~~ \langle \frac{\alpha_s}{\pi} G^2 \rangle_n= 
\langle \frac{\alpha_s}{\pi} 
G^2 \rangle_0 - \frac{8}{9} m_N^{(0)} \cdot n_{_B} \cdot B_1(x).
\end{eqnarray}
Here $\hat{m}$(1 GeV)$=(7 \pm2) $ MeV is the average value of the 
current quark masses of $u$ and $d$ quarks.  
The parameters we use are $\Sigma_{\pi N}=(45 \pm 7)$ MeV,
 $\langle 
\frac{\alpha_s }{\pi} G^2 \rangle_0=(350 {\rm MeV})^4$, 
 $\langle 
\bar{u}
u \rangle_0=(-230$MeV$)^3$\cite{HL92} and $m_N^{(0)} = 770-830$ MeV 
 with  $y \equiv 2 
\langle N| \bar{s} s |N \rangle/( \langle N| \bar{u} u |N \rangle 
 +\langle N| \bar{d} d |N \rangle ) = 0.22-0.12$ \cite{HL92,HK92}.
  Non-leading  $p_f$
corrections are  contained in $B_1(x)$ defined as 
\begin{eqnarray}
\label{b1}
 B_1(x)=\frac{3}{2x^3 }(
x \sqrt{1+x^2}-{\rm ln}[x+\sqrt{1+x^2}]) \simeq 1 - {3 \over 10} x^2
 + {9 \over 56} x^4 \cdot \cdot \cdot ,
\end{eqnarray} 
 where $x=p_f/m_N$
and  $n_{_B}=\gamma  p_f^3/(6\pi^2)$.  
For nuclear matter density $n_0=0.17 
 /fm^3$ ($x =  0.27$), 
 $x$ dependence of $B_1(x)$  can be safely neglected.
 Note also that  $\langle \bar{d}d \rangle_n
 =  \langle \bar{u}u \rangle_n $,  since
 we are considering the symmetric nuclear matter (N=Z).

As for the four quark condensate in medium,
  there are no experimental data
available yet.  Thus we use a simple mean field approximation
 in nuclear matter \cite{HL92};
\begin{eqnarray}
\label{saturation}
\langle (\bar{q} \gamma_{\mu} \gamma_5 \lambda^a q)^2 \rangle_n
 \simeq 
 - \langle (\bar{q} \gamma_{\mu} \lambda^a q)^2 \rangle_n
  \simeq  {16 \over 9} [
\langle (\bar{q} q)^2 \rangle_0 +2 \langle \bar{q} q \rangle_0
 \langle N |\bar{q} q |N \rangle \cdot n_{_B} \cdot  B_1(x)] \ \ ,
 \end{eqnarray}
 As for  $\langle (\bar{q} q)^2 \rangle_0$
 at 1 GeV scale, we will use the canonical value
  $(-281 {\rm MeV})^6$  \cite{SVZ} with
  $\alpha_s(1{\rm GeV}) \simeq 0.36$.
 This number, which  is substantially larger than the
 current algebra value $\langle \bar{q} q \rangle_0
 \simeq (-230 \ {\rm MeV})^3$ at 1 GeV, 
 should be considered as an effective one containing
  non-leading $1/N_c$ contributions.
  As for $\langle \bar{q} q \rangle_0$ in the second term of 
 (\ref{saturation}),
 it is not clear whether one should use
  $(-230 {\rm MeV})^3$ or $(-281 {\rm MeV})^3$.
 (Note that the latter number was used in
 \cite{HL92}.)
 Taking into account such ambiguity as well as the 
 ``experimental'' errors
  of $\Sigma_{\pi N}$ and $\hat{m}$,
 we  adopt $\langle \bar{q} q \rangle_0 \langle N \mid \bar{q} q 
\mid N  \rangle = (-256 {\rm MeV})^3 \cdot (45/14) \cdot
 (1 \pm 0.368)$
 as a standard value to be used in (\ref{saturation}).

The twist-2 quark bilinear operators with dimension 4 and 6 have non-
vanishing matrix elements in the medium.  Their 
 nucleon matrix elements are related to the parton 
distribution function in DIS as,

\begin{eqnarray}
  \left. 
 \langle N(p) |{\cal ST} \bar{q} \gamma_{\mu_1} D_{\mu_2} \cdot
 \cdot \cdot D_{\mu_n}q  |N(p) \rangle
  \right| _{\mu^2}=(-i)^{n-1} A_{n-1}^q(\mu^2) T_{\mu_1 \cdot 
 \cdot \cdot \mu_n}, \nonumber \\ [12pt]
A_{n-1}^q(\mu^2)=2 \int_0^1 dx x^{n-1} [q(x,\mu^2)+\bar{q}(x,\mu^2)].
\end{eqnarray}
Here  $ T_{\mu_1 \cdot \cdot \cdot \mu_n} \equiv [p_{\mu_1} 
\cdot \cdot \cdot p_{\mu_n}- ({\rm trace~~ terms})]/2m_N$ 
 and ${\cal ST}$ makes the operators symmetric 
and traceless.  For the  parton distribution function
 $q(x,\mu^2)$, we take 
the LO scheme in ref.\cite{GRV90} at 
the scale $\mu^2 = 1$ GeV which is close 
to the relevant Borel mass. Then we obtain  $A_1^{u+d} \simeq  0.9$,
 $A_3^{u+d} \simeq 0.12$ at 1 GeV.
 The fermi-motion corrections are contained in  
$B_2(x)=\sqrt{1+x^2}$ and $B_3(x)=\sqrt{1+x^2}(1+\frac{8}{5}x^2)$
 for spin 2 and spin 4 operators respectively. The deviation of
 $B_{2,3}(x)$
 from 1 is again small at nuclear matter density.

In reference 
\cite{HL92}, all the operators up to $d$=6 except for 
 relatively small  twist-4 spin-2 operators  are taken into account.
 In section IV, we will discuss the effect of the twist-4 operators
 to the result  of \cite{HL92}.

In the vacuum QSR, the spectral function
 (i.e. ${\rm Im} \Pi^R$
 in  eq.(\ref{dispersion}))
 is modeled with a  resonance pole and the 
continuum.   In the medium,  we have to add additional 
singularities below the lowest resonance pole  within the Fermi-gas 
approximation, which is called the Landau damping contribution 
\cite{BS}.
 For ${\bf q} \rightarrow 0$, it
 is calculable {\em exactly} 
 and behaves like  a pole at $\omega^2=0$
 (see Appendix A for the proof).
 In total,  the hadronic spectral function looks as
\begin{eqnarray}
\label{phen}
 8 \pi {\rm Im} \Pi^R(u > 0^-) & = & 
 \delta(u^2) \rho_{sc}+F \delta(u^2-m_V^2)
+(1+\frac{\alpha_s}{\pi})\theta(u^2-S_0) \\ \nonumber
 & \equiv & \rho_{had.}(u^2) ,
\end{eqnarray}
with $\rho_{sc} = 2 \pi^2 n_B /\sqrt{p_f^2 + m_N^2} \simeq
 2 \pi^2 n_B /m_N$.  $m_V$, $F$ and $S_0$ are the three
 phenomenological
 parameters to be determined by the sum rules.

Matching the OPE side 
 and the phenomenological side via the dispersion relation 
in the asymptotic region $\omega^2 \rightarrow - \infty$,
 we can   relate the resonance  parameters to the density 
dependent condensates.  
 There are two major procedure for this matching, namely the 
 finite energy sum rules (FESR) \cite{PIV}
 and the Borel sum rules (BSR) \cite{SVZ} which
 are  summarized as the following forms:
\begin{eqnarray}
\label{sumrules}
\int_0^{\infty}& ds\ W(s)& \ [\rho_{had.}(s) - \rho_{_{OPE}}(s) ]
  =0 ,\\ 
& W(s) & = \left\{ \begin{array}{ll}
                    s^n \ 
 \theta(S_0 -s) & \ \ \ \ \ \ \ \ \  ({\rm FESR}), \\ 
\nonumber
                    e^{-s/M^2} & \ \ \ \ \ \ \ \ \ ({\rm BSR}).
                   \end{array}     \right.
\end{eqnarray}
Here the spectral function $\rho_{had.}(s)$ stands for
 eq.(\ref{phen}).
 $\rho_{_{OPE}}(s)$ is a hypothetical imaginary part of
 $\Pi^R$ which, through
 the dispersion relation (\ref{dispersion}), reproduces 
eq.(\ref{ope}).
   For more details on (\ref{sumrules})
 and the explicit form of  $\rho_{_{OPE}}$, see section 2 of
  ref.\cite{HKL93}.

\vspace{0.8cm}

\noindent{\bf FESR and BSR  for $\Pi^R(\omega^2)$}

\vspace{0.5cm}

First, for the qualitative argument, let us  write down the FESR
 for the  rho (omega) meson in the chiral limit. 
 This can be easily obtained by 
 taking the first three moments $n=0,1,2$ in (\ref{sumrules}):

\begin{eqnarray}
\label{FESR}
F-S_0(1+\frac{\alpha_s}{\pi}) & = & -2\pi^2 m_{_N}^{-1} \cdot n_{_B} 
 \ \ \ \ \ \ \ \ \ (n=0),  \nonumber 
 \\ [12pt]
Fm_{_V}^2-\frac{S_0^2}{2} (1+\frac{\alpha_s}{\pi}) & = &
 -{\cal Q}_4-2 \pi^2 A_1^{u+d} 
m_{_N} n_{_B}  \equiv  - \tilde{{\cal Q}}_4 \ \ \ \ \ \ \ (n=1),
  \nonumber \\ [12pt]
Fm_{_V}^4-\frac{S_0^3}{3} (1+\frac{\alpha_s}{\pi}) & = & -{\cal Q}_6-
 \frac{10}{3} \pi^2 A_3^{u+d} m_{_N}^3 n_{_B}  \equiv  -
 \tilde{{\cal Q}}_6 \ \ \ \ \ \ \ (n=2),
\end{eqnarray}
where ${\cal Q}_4 \ ({\cal Q}_6)$
 is $\frac{\pi^2}{3} \langle {\alpha_s \over \pi}G^2
\rangle_n \
 (\frac{896}{81}\pi^3 \langle \alpha_s (\bar{q} q)^2\rangle_n)$ 
taken up to linear 
in $n_{_B}$. 
 Using the three relations above, we can determine the three 
phenomenological parameters $F,S_0$ and $m_V$ or equivalently,
 the changes  from the vacuum values $\delta F, \delta S_0$ 
and $\delta m_{_V}$. Important density dependence
 comes from $A_1^{u+d} n_{_B}$ and the 
 the 4-quark condensate ${\cal Q}_6$, which can be shown by solving
 eq.(\ref{FESR}) numerically.

Although both FESR and BSR give the same qualitative result,
 BSR is more reliable for the quantitative estimate since
 it is rather insensitive to the assumption on the continuum.
The rho (omega)  meson mass in the BSR is given as
\begin{eqnarray}
\label{sumHL}
 { m_{_V}^2 \over M^2}  = { 
 (1+{\alpha_s \over \pi}) \left(1-e^{-S_0/M^2}
 (1+ { S_0 \over M^2}) \right)- {1 \over M^4}\tilde{{\cal Q}}_4
  + {1 \over M^6} \tilde{{\cal Q}}_6 \over
  (1+{\alpha_s \over \pi}) \left(1-e^{-S_0/M^2}
  \right)+ {1 \over M^4} \tilde{{\cal Q}}_4  - {1 \over2  M^6}
 \tilde{{\cal Q}}_6  -\rho_{sc} }.
\end{eqnarray}
 In Fig.1, the Borel curve ($m_{_V}-M^2$ relation) is shown
 for different baryon densities.  The continuum threshold $S_0$
 is chosen to make the Borel curve as flat as possible in 
 the Borel window $M_{min}^2 < M^2 < M_{max}^2$ at given density.
 We take density-independent window $M_{min}^2=0.41 {\rm GeV}^2$
 and $M_{max}^2=1.30 {\rm GeV}^2$ in our analyses.  More general 
procedure with 
 density-dependent window (see e.g.  section 4 of ref.\cite{HKL93})
 does not change the results quantitatively.

\vspace{0.2cm}

\centerline{\fbox{\bf Fig.1}}

\vspace{0.2cm}

  By making a linear fit using the values at $n=n_0$ and $n=n_{_B}$,
 we get
\begin{eqnarray}
\label{mass-shift}
{m_{_V}(n_{_B}) \over m_{_V}(0)} & = & 1- (0.16 \pm 0.06) 
{n_{_B} \over n_0}, \\
\label{threshold-shift}
\sqrt{{S_0(n_{_B}) \over S_0(0)}} & = & 1- (0.15 \pm 0.05)
 {n_{_B} \over n_0}, \\
\label{F-shift}
{F(n_{_B}) \over F(0)} & = & 1- (0.24 \pm 0.07) {n_{_B} \over n_0}, 
\end{eqnarray}
 These values are slightly different from our previous
 ones in \cite{HL92} where  the  uncertainty
  discussed below 
 (\ref{saturation}) is not taken into account.
 For the decreasing rho (omega)  mass,
 the twist 2 and the scalar 
 matrix elements are equally important.

\section{Mass shift and the scattering length}

 In this section, we will first summarize the claims given
  explicitly or
 implicitly in \cite{Koike93}. 
 Although they look plausible in the first look,
 every statement is invalid as we will show
 in section III.A-C.

\vspace{0.5cm}

\noindent
{\bf (A)} Koike
 starts with eq.(\ref{fermicor}) and makes a low density 
approximation for the second term 
\begin{eqnarray}
\label{A1}
 \gamma \int^{p_f}  {d^3 p \over (2\pi)^3 2E_N} T^R_{\mu \nu}
 (\omega, {\bf q} \mid {\bf p})
 \rightarrow n_{_B}\ {T^R_{\mu \nu}(\omega,{\bf q} \mid {\bf 0})
 \over 2m_{_N}} .
\end{eqnarray}
This corresponds to the assumption that 
  all the nucleons in nuclear medium are at rest
 (${\bf p}=0$).
 If one further takes the kinematics  ${\bf q}=0$ and
 $\omega \simeq m_{_V}$, 
 $T^R \equiv T^R_{\mu \mu}(\omega, {\bf q}=0 \mid {\bf 0})$
is written by the
 $V-N$ scattering length as
\begin{eqnarray}
\label{laurant}
 T^R \simeq
  { 3Fm_{_V}^2 \over 8 \pi^2} ~~ {24 \pi (m_{_N}+m_{_V}) a_{_{VN}} 
 \over (\omega^2-m_{_V}^2)^2}+ {\rm R(\omega^2)} ,
\end{eqnarray}
where  $a_{_{VN}}=(a_{1/2}+2 a_{3/2})/3$ with 
$a_{1/2}$ and $ a_{3/2}$ being the $V-N$ scattering length in 
the spin-1/2 and spin-3/2 channel respectively. 
${\cal R}(\omega^2)$ is the
 term less singular than the leading double-pole term in the 
 Laurent expansion around $\omega^2 =m_V^2$.

By substituting these expressions to eq.(\ref{fermicor}) 
and take the  leading term in eq.(\ref{laurant}), one arrives at
the formula
\begin{eqnarray}
\label{mass}
{1 \over 3} \Pi^R_{\mu \mu}(\omega \simeq m_{_V},{\bf q}=0;n_{_B})
 & \simeq &  { F m_V^2 \over 8 \pi^2}  \left(
 {1 \over \omega^2-m_{_V}^2} + 
 {12 \pi a_{_{VN}}(m_{_N}+m_{_V})/m_{_N}
 \over (\omega^2-m_{_V}^2)^2} \ n_{_B} 
\right),
 \nonumber \\ [12pt]
 & & \propto   {1 \over \omega^2-(m_{_V} + \delta m_{_V})^2}
\end{eqnarray}
with
\begin{eqnarray}
\label{deltam}
 \delta m_{_V} = 6 \pi  {m_{_N}+m_{_V} 
\over m_{_N} m_{_V}}\cdot  a_{_{VN}} \cdot n_{_B}.
\end{eqnarray}
Hence the positive (negative)
 scattering length gives an increasing (decreasing) mass in the
 medium.

\vspace{0.5cm}

\noindent
{\bf (B)} To estimate the magnitude and sign of the scattering length
 $a_{_{VN}}$ in (\ref{deltam}),
 Koike formulated  QCD sum rules
 for $T^R$.
 He starts with the unsubtracted dispersion relation;
\begin{eqnarray}
\label{Tdispersion}
{\rm Re}\  T^R (\omega^2) = {1 \over \pi } 
  \int_0^{\infty} du^2 {{\rm Im}T^R(u) \over u^2 - \omega^2}.
\end{eqnarray}
 The OPE for $T^R$ is expanded up to $O(1/Q^4)$ as 
\begin{eqnarray}
\label{koikeT1}
 {\rm Re}\  T^R_{_{OPE}}(Q^2) ={1 \over 8 \pi^2} \left( 
 {c_1 \over Q^2} - {c_2 \over Q^4}  \right),
\end{eqnarray}
where $c_{1,2}/2m_{_N} 
 \equiv d\tilde{{\cal Q}}_{4,6}/dn_{_B}$ at $n_{_B}$=0.  
The absence of the logarithmic term
 in (\ref{koikeT1}) indicates that  subtraction is not necessary
 in (\ref{Tdispersion}). Motivated by eq.(\ref{laurant}),
 the imaginary part is parameterized as
\begin{eqnarray}
\label{koikeT3}
8 \pi {\rm Im} T^R(u >0) =
 b_1 \delta'(u^2 - m_{_V}(0)^2) 
 + b_2 \delta (u^2 - m_{_V}(0)^2)
 + b_3 \delta (u^2 -S_0(0)),
\end{eqnarray}
with three unknowns  $b_{1,2,3}$ and  known 
 vacuum parameters $m_V(0)$ and $S_0(0)$.
 (Note that $b_1=F m_{_V}^2 24 \pi(m_{_N}+m_{_V}) a_{_{VN}}$.)
 Above parametrization is equivalent to taking the following
 ansatz for the real part
\begin{eqnarray}
\label{koikeT2}
 {\rm Re}\ T^R_{had.}(Q^2) =
 {b_1 \over (m_{_V}(0)^2+Q^2)^2} + {b_2 \over m_{_V}(0)^2 + Q^2}
              + {b_3 \over S_0(0) + Q^2}.
\end{eqnarray}
 By constructing a borel sum rule using (\ref{Tdispersion})$-$
 (\ref{koikeT3}), Koike  obtains positive scattering length 
$a_{\rho(\omega)} \simeq 0.14 \ (0.11)$ fm, from which
  he concluded that the mass shift $\delta m_{_V}$ in 
 eq.(\ref{deltam}) must be positive.

\vspace{0.5cm}

\noindent
{\bf (C)} Koike further claims that  the above procedure (A)+(B)
 is equivalent with
 doing the medium sum rule for $\Pi^R_{\mu \mu}$ but not for
  $\Pi^R$. (Note 
  that $\Pi^R_{\mu \mu} = -3 \omega^2 \Pi^R$ when ${\bf q}=0$).
 This can be seen as follows.  The 
dispersion relation in medium for $\Pi^R_{\mu \mu}$ reads
\begin{eqnarray}
\label{dispersion2}
  {\rm Re} \Pi^R_{\mu \mu} (\omega^2) =  {1 \over \pi}
 {\rm P} \int_0^{\infty} du^2
{ {\rm Im} [ \Pi^R_{\mu \mu}(u)] \over u^2-\omega^2} +
 ({\rm subtraction}).
\end{eqnarray}
 If one adopts eq. (\ref{phen}) for ${\rm Im} \Pi^R$  
  and uses the relation $\Pi^R_{\mu \mu} = -3 \omega^2 \Pi^R$,
 one obtains
\begin{eqnarray}
\label{phen2}
(-{1 \over 3}) 
 8 \pi {\rm Im}[ \Pi^R_{\mu \mu}(u)]
 & = & Fm_V^2 \delta(u^2-m_V^2)
+(1+\frac{\alpha_s}{\pi}) u^2 \theta(u^2-S_0), \nonumber \\
   & = & u^2 \rho_{had.}(u^2). 
\end{eqnarray}
 Since $u^2 \delta(u^2) =0$,
 the Landau damping term in (\ref{phen})
 does not arise in (\ref{phen2}).

Expansion of the l.h.s. of (\ref{dispersion2}) 
in terms of $n_B$ gives
 [$n_B$ independent term]  + [$n_B \times $ eq.(\ref{koikeT1})],
 while the same expansion of
 (\ref{phen2}) gives 
 [$n_B$ independent term]  + [$n_B \times $ eq.(\ref{koikeT3})].
 The latter is obtained simply by writing 
 $m_{_V} = m_{_V}(0) + \delta m_{_V}$,
 $S_0 = S_0(0) + \delta S_0$, $F=F(0)+\delta F$,
  expanding (\ref{phen2}) up to linear in
 $\delta m_{_V}, \delta F$ and $\delta S_0$ and doing the
following identification.
\begin{eqnarray}
\label{id}
b_1/2m_N=-Fm_{_V}^2 \delta m_{_V}^2,~~~b_2/2m_N=m_{_V}^2 \delta F + F
\delta m_{_V}^2,~~~ b_3/2m_N=-S_0 \delta S_0.
\end{eqnarray} 
 
This means that  
 the sum rule for $T^R$ eq.(\ref{Tdispersion})
is equivalent to the linear density
  part of the  sum rule for 
 $\Pi^R_{\mu \mu}$ eq.(\ref{dispersion2}). 
 Assuming that his procedure (A)+(B) is right,
 Koike thus concluded that
 (i) the  medium QSR using $\Pi^R_{\mu \mu}$ must give the
  increasing vector-meson mass, and (ii) 
   the result of the medium sum rule using $\Pi^R$ in 
 \cite{HL92} must be wrong.

 In the following, we will discuss that each of the above arguments 
 is invalid.
 The subsection numbers III.A, III.B and III.C 
  in the following correspond to
 the statements (A), (B) and (C) in the above, respectively.

\subsection{Mass shift and the scattering length}       

 First of all, let us show that eq.(\ref{A1}) is a useless 
 approximation around nuclear matter density.  
 In  eq.(\ref{A1}), the motion of nucleons  and
  the Pauli exclusion principle in nuclear matter are completely
 neglected.
 Such approximation is
  valid only when (i)  the nucleon density is extremely low, or
 (ii) $T^R_{\mu \nu}(\omega, {\bf q} =0 \mid {\bf p}) $ is almost
 constant as a function of ${\bf p}$ in the interval
  $0 < \mid {\bf p \mid} < p_f$.

 Since we are not interested in the case (i), let us concentrate on
 (ii) and see whether (ii) is plausible or not.
 At nuclear matter density, the fermi momentum is sizable 
 $p_f \simeq 270 $ MeV. Thus we should
 consider e.g. the $\rho-N$ scattering 
 from $\sqrt{s} =  m_{\rho} + m_N = 1709$ MeV
 through $\sqrt{s} = 
((m_{\rho} + \sqrt{m_N^2 + p_f^2})^2-p_f^2)^{1/2} 
= 1726$ MeV.  In this interval,
  there are at least two s-channel resonances
 $N(1710)$ and $N(1720)$ and also there are two nearby resonances 
 just below the threshold $N(1700)$ and $\Delta(1700)$  \cite{PDG}.
 They are all possible to have coupling with $\rho  N$.
 This means that $T^R_{\mu \nu}$
  has a rapid variation as a function of $\mid {\bf p} \mid$
 between $\mid {\bf p} \mid =0$ and $\mid {\bf p} \mid =p_f$
 due to the effect of these s-channel resonances  and
 it is impossible to approximate it by the threshold
 value (i.e. the $V-N$ scattering length)
 $T^R(\omega=m_V,{\bf q}=0 \mid {\bf p}=0)$. See Fig.2 for a
 schematic illustration of the $s-$channel contributions
 \cite{tchan}.

\vspace{0.2cm}

\centerline{\fbox{\bf Fig.2}}

\vspace{0.2cm}

   What one can expect at best is the approximate linear density
formula written in terms of the average of $T^R_{\mu \nu}$ 
 in the region $0< \mid {\bf p} \mid <p_f$:
\begin{eqnarray}
\label{AA1}
 \gamma \int^{p_f}  {d^3 p \over (2\pi)^3 2E_N} T^R_{\mu \nu}
 (\omega, {\bf q}=0  \mid {\bf p})
 \simeq  n_B \ \left< {T^R_{\mu \nu}(\omega,{\bf q}=0 \mid {\bf p})
 \over 2E_N} \right>  ,
\end{eqnarray}
where $\langle \cdot \rangle$
 stands for the average over the above momentum interval.
 
 It is easy to see what is wrong in (\ref{A1}) in physical terms:
 the rho-meson at rest (${\bf q}=0$)
  in nuclear matter will suffer  the scattering
 from the nucleons having various three-momentum ${\bf p}$
 in the interval $0 < \mid {\bf p} \mid <  p_f$.
 The net effect should be  the averaged strength of the scattering 
 and not the scattering length defined at ${\bf p}=0$. 
  This is particularly so when $T^R_{\mu \nu} $
 has a rapid ${\bf p}$ dependence.

The above point is well known for the nucleon in nuclear medium.
 The optical potential for the nucleon at rest in
  nuclear matter cannot be
 approximated by the N-N scattering length multiplied by the 
nuclear density.
 In fact, the N-N forward scattering amplitude $T_{NN}(p)$
 has a huge momentum dependence due to the 
  the deuteron state and the ``almost'' bound state
  near the  threshold.
   The relevant quantity for the nucleon optical potential 
 is not  the scattering
 length but the averaged scattering amplitude
 in the interval $0 < \mid {\bf p} \mid < p_f$. 
  This point was recently emphasized by
  Furnstahl and one of the present authors \cite{FH93}:
  They  criticise the paper by  Kondo and Morimatsu \cite{KM93}
  who use a similar approximation with (\ref{A1})  
  to analyse
  the nucleon in nuclear medium.

  It is now clear that
  the mass shift and the scattering length cannot be related directly
 at nuclear matter density.
 In the approach of ref.\cite{HL92}, neither eq.(\ref{A1}) nor
 eq.(\ref{AA1})
 are adopted, thus one does not suffer from this problem. 

 \subsection{Is scattering length calculable in QSR?}       

Since eq.(\ref{A1}) and hence 
 eq.(\ref{deltam}) are not valid around nuclear matter density,
  it is useless to
 relate the scattering length with the mass shift. 
 Nevertheless, the $V-N$ scattering length itself could be an 
 interesting physical quantity to be calculated in QSR.
 We will show, however, that the method in ref.\cite{Koike93} to
 estimate $a_{_{VN}}$  is erroneous.
 
 Let us  look at (\ref{koikeT1}) and (\ref{koikeT2}) and
 compare them at $Q^2 \rightarrow \infty$,
 which corresponds to the FESR  for $T^R$.
  One immediately realizes that  only
 two independent equations can be  obtained  
\begin{eqnarray}
\label{twoeq}
 b_2+ b_3 = c_1, \nonumber \\
   b_1 -m_{_V}(0)^2\ b_2 - S_0(0)\ b_3 =- c_2,
\end{eqnarray}
whereas one needs three equations to solve
 $b_{1,2,3}$.  This happens because the OPE 
 is calculated only up to
 $O(1/Q^4)$  in (\ref{koikeT1}).  (\ref{twoeq}) clearly shows that
 it is impossible to predict
 the scattering length $a_{VN}$ (which is proportional to $
b_1$).

 In QCD sum rules, one should always check 
 that the number of phenomenological parameters to be determined 
 is equal or smaller than the number of OPE terms, otherwise
 sum rules are not closed and cannot give predictions.
 The FESR provides a useful tool to do this
 consistency check.

 If one tries to make Borel analyses without the consistency 
as has been done in ref.\cite{Koike93}, 
 one  simply obtains a 
 fake result and does not get any stability
 of the Borel curve.
  By blindly doing a Borel
sum rule 
without looking for a stability region, Koike
 obtains a positive number for $b_1$, in which  it  is implicitly
  assumed that the contribution from dimension 8 operators is  
zero.  In terms of finite energy sum rule, this amounts to the
following condition,
\begin{eqnarray}
\label{addi}
2m_{_V}(0)^2\ b_1-m_{_V}(0)^4\  b_2 +S_0(0)^2\  b_3=0.
\end{eqnarray}
With this assumption and eq.(\ref{twoeq}), 
 we have three equations and the unknown
constants can be determined.  Using the fact that $c_1 \sim 0$, 
$c_2 >0$ and $S_0 > m_{_V}^2$, one finds a positive value for $b_1$.
 However, the assumption eq.(\ref{addi}) has no ground
 and we do expect non-negligible contribution from the 
 dimension 8 operators.
 Since it is technically very hard to calculate these dim. 8
 condensates ( $1/Q^6$ terms
 of $T^R$), it is almost hopeless to get reliable
 $a_{_{VN}}$ in QSR.

\subsection{Use of QSR for $\Pi^R_{\mu \mu}$}

Here, we will explicitly demonstrate that the sum rules for
 $\Pi^R_{\mu \mu}$ recommended in \cite{Koike93} does not work at 
 all.
 Let's first start with the FESR for $\Pi^R_{\mu \mu}$;
\begin{eqnarray}
\label{dispPR}
\int_0^{S_0} ds\ s^n \ [ {\rm Im} \Pi^R_{\mu \mu}(s)_{had.}
 - {\rm Im} \Pi^R_{\mu \mu}(s)_{_{OPE}}] = 0 \ \ \
  (n=0,1,2, \cdot \cdot \cdot ).
\end{eqnarray}
 Then one immediately finds that only two relations 
 corresponding to  $n=0,1$ are
  obtained, and they turn out to be  
  equivalent with the second and third relations in eq.(\ref{FESR}).
 (Note that n=2 in (\ref{dispPR}) cannot gives another condition 
because OPE is calculated only up to dimension 6 operators.)
\begin{eqnarray}
\label{FESR2}
Fm_V^2-\frac{S_0^2}{2} (1+\frac{\alpha_s}{\pi}) & = & 
 -\tilde{\cal Q}_4,
 \ \ \ \ \ (n=0) \nonumber \\ [12pt]
Fm_V^4-\frac{S_0^3}{3} (1+\frac{\alpha_s}{\pi}) & = & 
 -\tilde{\cal Q}_6
 \ \ \ \ \ (n=1).
\end{eqnarray}
There are three unknowns $F,m_{_V},S_0$, while only two relations 
 are available.  
 Thus unless one introduces extra assumption,
 it is impossible to solve for three {\em even in the vacuum}.
 This is exactly the same problem which we have discussed
 in sec.III.B.

One should also note that the missing condition is the duality 
 relation
 for the spectral density:
\begin{eqnarray}
\label{duality}
\int_0^{S_0} ds \  [ \rho_{had.}(s)-\rho_{_{OPE}}(s) ]=0,
\end{eqnarray}
with  $\rho_{had.}$ and $\rho_{_{OPE}}$ being the spectral densities
 for $\Pi^R$.
  This local  duality is the cornerstone of vacuum
 QSR and  holds also in medium since there are no dimension 2 
operators in OPE.
 If one wants to get reliable result from FESR, 
   One has either to work out the OPE up to 
dimension 8 operators (which is a formidable task) or to
 start with $\Pi^R$ as in ref.\cite{HL92}.

 Here one may ask that ``why not take $n=-1$ moment in (\ref{dispPR})
 to obtain another relation?''.
 Such procedure, however,  introduces 
 an  ambiguity at $s=0$, since one can add any function
  proportional to   
 $s \delta(s)$ to ${\rm Im} \Pi^{R}_{\mu \mu}$
 which does not modify $n=0,1$ sum rules but modifies
 $n=-1$ sum rule. 
 If one tries to remove this ambiguity,
 it is necessary to start with
 $\Pi^R$ and to evaluate the Landau damping term in (\ref{phen})
which is exactly calculable
 as we mentioned \cite{further}.

Let us now turn to the BSR for  $\Pi^R_{\mu \mu}$:
\begin{eqnarray}
\label{sumKO}
 { m_{_V}^2 \over M^2}  = { 
 2(1+{\alpha_s \over \pi}) \left(1-e^{-S_0/M^2}
 (1+ { S_0 \over M^2}+{S_0 \over 2M^2}) \right) 
-  {1 \over M^6} \tilde{{\cal Q}}_6 \over
  (1+{\alpha_s \over \pi}) \left(1-e^{-S_0/M^2}(1+
  { S_0 \over M^2})
  \right)- {1 \over M^4} \tilde{{\cal Q}}_4  + {1 \over  M^6}
 \tilde{{\cal Q}}_6  }.
\end{eqnarray}
 Eq.(\ref{sumHL}) and eq.(\ref{sumKO}) in this paper correspond
 to eq.(13) and eq.(14) in \cite{Koike93} respectively.
 Since $ \tilde{{\cal Q}}_6$ enters with opposite sign in the r.h.s.
 of eq.(\ref{sumHL}) and eq.(\ref{sumKO}) and 
 $ \tilde{{\cal Q}}_6$ decreases in medium,
  Koike simply concluded that $m_{_V}$ decreases in
  (\ref{sumHL}), while it increases in (\ref{sumKO}).
  This conclusion is too naive:
  in fact, $S_0$ in the r.h.s. of these equations is also
 density dependent, which can change such naive expectation.
  density dependence of $S_0$ can be in principle determined by
 the Borel stability procedure we have discussed in section II.
 
 In order to see whether this procedure works or not
 for eq.(\ref{sumKO}), we have shown the Borel curves for
  different values of $S_0$ in Fig. 3(a) (at  zero density) 
 and in Fig.3 (b) (at nuclear matter density).
  Fig. 3(a) shows that the Borel curve does not have any
 plateau in the relevant range of $M^2$ (say
 $0.41 < M^2 < 1.30$), which implies that one cannot 
 determine  $S_0$ and hence $m_{_V}$ even in the vacuum.
  The situation is the same at the nuclear matter density as is shown
 in Fig.3(b). Again, one cannot determine
 $S_0(n_0)$ by the Borel stability method and hence $m_{_V}(n_0)$,
 which implies that there is no hope to determine the mass shift
 at finite density.
 If one sticks to  a specific value of 
 $S_0$ (say $2.0 {\rm GeV}^2$) and uses
 it at any density, one finds that $m_{_V}(n_{_B})/m_{_V}(0) >1$
 for given $M^2$ from Fig.3. This is equivalent with the
  ``naive'' (and wrong) argument by Koike.
  The correct procedure is
  to compare $m_{_V}(n_{_B})$ (calculated
 with $S_0(n_{_B})$) and $m_{_V}(0)$ (calculated with $S_0(0)$).

\vspace{0.2cm}

\centerline{\fbox{\bf Fig.3}}

\vspace{0.2cm}

  The ``bad'' Borel curves in Fig. 3
 is quite in contrast to the ``good'' Borel curves for $\Pi^R$
 in Fig.1. The latter shows beautiful stability in the vacuum as
 well as in the medium, which makes one  possible to
 determine $S_0(n_{_B})$ at each density and hence $m_{_V}(n_{_B})$.

The reason of the failure of the BSR for $\Pi^R_{\mu \mu}$ is 
  twofold. 
  Firstly, the higher dimensional operators in OPE is rather
 important for  $\Pi^R_{\mu \mu}$ sum rules.
 We have already seen this  in (\ref{FESR2})
 where $n=2$ sum rule can be obtained only when one 
 has  dim. 8 operators in OPE.  In BSR, the lack of 
 the information of dim. 8 operators arises as a 
 instability of the Borel curve at low $M$ region.
 Inclusion of the dim. 8 operator would make the curve more flat.
 Secondly, the continuum contribution is more
 important in $\Pi^R_{\mu \mu}$ than
  in $\Pi^R$, since the spectral function is increasing linearly
 in the former case.  This makes the prediction of the 
 resonance parameters  less reliable in the former.

 Let us summarize here the lessons we learned in subsections III.A,
 III.B and III.C.  Firstly, the mass shift and the scattering length
 does not have direct relation in nuclear matter due to the
 momentum dependence of the $V-N$ forward scattering amplitude.
  Secondly, sum rules for the $V-N$ scattering amplitude
 cannot predict the $V-N$ scattering length without
 dimension 8 operators in OPE.
 Thirdly, sum rules for $\Pi^R_{\mu \mu}$
   does not work at all even in the vacuum
 without  dimension 8 operators in OPE.
  Thus all the claims  in ref.\cite{Koike93} are shown to be 
  erroneous. Also, only the consistent QSR in medium
    currently available is the one starting from $\Pi^R$ 
 given in ref.\cite{HL92}.

\vspace{1cm}

\section{Several comments}

\noindent{\bf  Full twist 4 calculation}

There are three 
kinds of twist-4 spin-2 operators  contributing to 
the $\rho,\omega$ sum rules.   
At present, their nucleon matrix 
elements are not known.  However, the values of 
two different combinations in
  the transverse and longitudinal structure functions
 of the nucleon have 
been obtained by two of us \cite{CHKL93,Lee94} by analyzing the 
recent DIS data at CERN and SLAC.  
Let us  further make the following assumption\cite{CHKL93};

\begin{eqnarray}
 {\langle \bar{d} \Gamma_\mu \Delta_\nu d \rangle \over
  \langle \bar{u} \Gamma_\mu \Delta_\nu u \rangle} =
 {\langle \bar{d} \gamma_\mu D_\nu d \rangle \over
  \langle \bar{u} \gamma_\mu D_\nu u \rangle} \equiv \beta,
\end{eqnarray}
where $\Gamma_\mu$ is some  gamma matrix and $\Delta_\mu$ an isospin 
singlet operator.  
  Then, it is possible to uniquely determine the nucleon matrix
element of twist-4 spin-2 operators  appearing in the 
 $\rho,\omega$ sum rule from  combinations of experimental values.
It gives the following contribution to eq.(\ref{Tmatrix}), 

\begin{eqnarray}
\label{twist4}
T^{R,\tau=4}_{\mu \mu}(\omega, {\bf q} | {\bf p})= 
 {-[ (q \cdot p)^2-\frac{1}{4}m_N^2 q^2 ] \over m_N Q^4} 
 \left(- (1+ \beta)
 (K_u^1+\frac{1}{4}K_u^2+ \frac{5}{8} K_u^g)+K_{ud} \pm 
  K_{ud} \right)
\end{eqnarray}

The $K_u^i$ $i=1,2,g$ are defined in ref.\cite{CHKL93} and $- (+)$ 
corresponds to the $\omega (\rho)$ case.
Choosing  $\beta=0.476$ as in ref.\cite{CHKL93} the value inside 
the large round bracket 
is $A^4=0.40$ (0.24) GeV$^2$ for the $\omega$ ($\rho$) meson.  

Now the effect of twist-4 matrix element can be estimated by 
 making the 
following substitution for the dimension six operators.
\begin{eqnarray}
({\cal Q}_6+\frac{10}{3} \pi^2 A_3^{u+d} m_{_N}^3 n_{_B})\rightarrow
({\cal Q}_6+\frac{10}{3} \pi^2 A_3^{u+d} 
 m_{_N}^3 n_{_B}+2\pi^2m_{_N} A^4 n_{_B})
\end{eqnarray}

The net effect of twist-2 + twist-4  is estimated to be
 2.36 (3.29)  times larger than the
 twist 2 effect alone in the rho (omega) channel.
 This could change the slope of the mass shift in
 (\ref{mass-shift}) from 0.16 to 0.10 (0.075) for the rho (omega)
 meson.
 Further investigation is necessary however to draw definite
 conclusion on the magnitude of the twist 4 effect.

\vspace{0.5cm}

\noindent{\bf Fermi momentum correction}

The small higher density effects coming from fermi-momentum 
 correction can 
be estimated by looking at  $B_i(x)$.  One easily finds that
 $B_1(0.27)=0.979, B_2(0.27)=1.036, B_3(0.27)=1.157$ and
$B_1(0.34)=0.967, B_2(0.34)=1.056, B_3(0.34)=1.251 $,
 where $x=0.27 (0.34)$ corresponds to
 nuclear matter (twice of nuclear matter) density. 
 Thus the effect can be safely neglected at  nuclear matter density.

\vspace{0.5cm}

\noindent{\bf Possible new structure in ${\rm Im} \Pi^R$}

It is possible that the density dependent change of the
 OPE side
 is balanced by  some new structure appearing in the
spectral density below the resonance mass.  In  QCD sum rule
approach, it has to be included by hand before matching to the OPE.
Such possibility has been examined by Asakawa and
Ko\cite{AK93} by redoing the medium QCD sum rules for the vector
meson including other  complex structure of the spectral
density in the nuclear medium induced by the
 $\pi, \Delta, N, \rho$ dynamics.   They found that even in that case
  the  vector meson mass has to decrease in order to be
consistent with the OPE side.

\vspace{0.5cm}

\noindent{\bf $\phi$ meson sum rule}

The formalism for calculating the change of $\phi$ meson sum
rule is the same as that of the $\rho$ and $\omega$.  However,
in the $\phi$ case, one must include the effect of the strange quark
mass in the OPE and this will introduce some basic difference
 \cite{HL92}.   In the $\rho,\omega$ sum rule, the 
 density dependence in OPE is 
 dominated by $(d,\tau)=(4,2)$ and 
 $(d,\tau)=(6,0)$ operators.
 However, in the case of the $\phi$ meson, the dimension 4
strange quark condensate $ \langle m_s \bar{s} s \rangle$
 is not suppressed  either by $1/4 \pi^2$ or the light current
quark mass and consequently dominates the OPE.
 In the medium, the change of this condensate, which comes from
the  the K-N sigma term, dominates the small changes in
other condensate and  introduce a non-negligible mass reduction
by $3 \sim 5$\% \cite{HL92}.  
It is amusing to compare this result with that of an effective model 
calculations\cite{CMK92} in which the K-N sigma term
  also induces a small reduction
of the $\phi$ mass. 

\vspace{1cm}

\section{Summary}

To estimate the mass shift of vector mesons in medium, 
we have carried out a detailed comparision between the 
 approach based on the
modification of the vacuum QSR \cite{HL92} and that based on the
scattering length \cite{Koike93}.
 We have shown that the latter approach
 is erroneous by the following reasons:

\noindent 
(i) The mass shift and the scattering length
 does not have direct relation in nuclear matter due to the
 momentum dependence of the $V-N$ forward scattering amplitude.

\noindent
(ii) 
 Sum rules for the $V-N$ scattering amplitude
 cannot predict the $V-N$ scattering length if
 dimension 8 operators in OPE are not included.
 
\noindent
(iii) 
 Sum rules for $\Pi^R_{\mu \mu}$
   have no predictive power  both in the vacuum and in the medium
 if  dimension 8 operators  are not included.

\acknowledgments

 The authors thank Dr. Y. Koike for stimulating discussions
 to clarify the issue, although he does not share our
 arguments.
 TH thanks Dr. K. Tanaka for useful comments 
 and Prof. M. Ichimura for discussions and useful 
  information on nucleons in nuclear matter. 
 TH and SHL  thank the Institute for Nuclear Theory at the 
University of Washington for its hospitality
 and the Department of Energy for partial support
 during the completion of this work. 
 The work of TH and HS was supported in part by the
 Grants-in-Aids of the Japanese Ministry of Education (No. 06102004).
 The work of SHL was supported by the Basic Science Research 
Institute  program of the Korean Ministry of Education 
through Grant No. BSRI-94-2425 and  
by KOSEF through the CTP at Seoul National University.

\newpage
\appendix
\section*{A}

Here, we derive the Landau damping term $\rho_{sc}$.  
In the Fermi gas approximation, the spectral density has two types 
of contribution.  The annihilation term, which is non-zero 
above the two 
particle threshold $\omega^2 > {\bf q}^2+4m_N^2$,  and the scattering
 term, which is non-zero in the space-like region
  $\omega^2< {\bf q}^2$. 
We are interested in the second term.   For finite ${\bf q}$, the 
spectral density, contributing to the longitudinal polarization,
  can be 
obtained by looking at the $00$ component of the 
imaginary part of eq.(\ref{correlator}). 
\begin{eqnarray}
\rho_l^s( \omega, {\bf q})={ {\rm Im} \Pi_{00} \over {\bf q}^2}
~~~ \stackrel{ {\bf q} \rightarrow 0}{\longrightarrow }
~~~ { {\rm Im} \Pi_{\mu \mu} \over -3 \omega^2} .
\end{eqnarray}
 
Looking at the spectral representation, it is easy to identify
 the following scattering contribution,
\begin{eqnarray}
\label{scat1}
\rho_l^s(\omega, {\bf q}) & = & { (2 \pi)^4  \over 4 {\bf q}^2} 
\int {d^3  k_1 \over (2\pi)^3 E_1}{ d^3  k_2 \over (2 \pi)^3 E_2 }
 |\langle N({\bf k}_1)|J_0| N({\bf k}_2) \rangle |^2  \\ \nonumber
 &  &  \ \ \ \ \ \ \ \ \ \ \ \ \ \ \ \ \ \ \times
\delta(\omega-E_1+E_2) \delta^3( {\bf q-k_1+k_2})
 \left[n_F(E_2)-n_F(E_1) \right]\ \ .
\end{eqnarray}
Here,  $E_i=\sqrt{ {\bf k}_i^2+m_N^2}$ (i=1,2), 
and the $n_F(E_i)= 
\theta ( \sqrt{ k_f^2+m_N^2}-E_i)$, where $ k_f$ is 
the fermi momentum.

In general, the nucleon expectation of the isospin current has
 two form 
factors.
\begin{eqnarray}
\label{form}
\langle N({\bf k_1})|J^a_\mu| N({\bf k_2}) \rangle= 
\bar{u}({\bf k_1}) {\tau^a \over 2} \left[
 F_1(q) \gamma_\mu +F_2(q) i \sigma_{\mu \nu} q^\nu 
 \right] \bar{u}({\bf k_2}) .
\end{eqnarray}

Substituting this into eq.(\ref{scat1}), one obtains
\begin{eqnarray}
\rho_l^s(\omega,{\bf q}) & = &
{\gamma \over 256 \pi^2} \int_v^\infty 
dx \left[(1-x^2)F_1^2(q)+x^2 (\omega^2 
 - {\bf q}^2) F_2^2(q) \right] \\ \nonumber
 & & \ \ \ \ \ \ \ \ \ \ \ \ \ \ \ \ \ \ 
\ \ \  \times \left(2 n_F \left( {|{\bf q}|x+\omega \over 2} \right) 
 - 2 n_F \left( {|{\bf q}|x-\omega \over 2} \right) \right)  
\theta({\bf q}^2-\omega^2),
\end{eqnarray}
where $v=[1-4m^2/(\omega^2-{\bf q}^2)]^{1/2}$.

Now, we want to take the limit $|{\bf q}|\rightarrow 0$.  
In this limit, the constraint  $0<\omega^2<{\bf q}^2 $ 
also forces $\omega$ to approach zero.  
Consequently, the 
contribution proportional to  $F_2$ vanishes, because it is
 multiplied by either  ${\bf q}^2$ or  $\omega^2$. 
 As for the other term 
proportional to $F_1(q)$,  the integral becomes increasingly 
large as 
 $| {\bf q}| \rightarrow 0$ such that the integrated quantity of 
 $\rho_l^s(\omega,{\bf q})$ within the phase 
space for $\omega$ remains finite. 

By integrating over this region with $| {\bf q}|$ finite 
and then taking the limit,  we find
\begin{eqnarray}
\label{scatf}
\lim_{|{\bf q }| \rightarrow 0} \int_0^{{\bf q}^2} d\omega^2   
 \rho_l^2(\omega,{\bf q})
= {\gamma \over 12} \int { d^3 p \over (2\pi)^3 2 E} n_F(E) v(3-v)
\equiv \rho_{sc}/8 \pi \ ,
\end{eqnarray}
so that $\rho_l^s(\omega,p)$ effectively becomes a delta function.  
Thus 
the final result is that the spectral density reduces to 

\begin{eqnarray}
\lim_{|{\bf q}| \rightarrow 0} \rho_l^s(\omega, {\bf q}) =
\delta(\omega^2) \rho_{sc}/8\pi \ \ .
\end{eqnarray}

It should be noted here that 
we did not make any approximation from
 (\ref{scat1})  through (\ref{scatf}), thus
 the result is exact. Also note that there arises no ambiguity
 from the nucleon form factor because
 $F_1(q=0)=1$.

\newpage
\centerline{\bf{Figure Captions}}

\vspace{0.5cm}

\begin{description}

\item{Fig.1:} Borel curve for  $m_{_V}(M^2)$  using $\Pi^R$.
 Solid, dashed, dash-dotted lines correspond
 to $n_{_B}/n_0 = 0,  1.0, 2.0$ respectively.
 $S_0(n_{_B})$ determined by the Borel stability method
 at each density is also shown in GeV$^2$ unit.
 The Borel window is chosen to be  
 $0.41 {\rm GeV}^2 < M^2 < 1.30 {\rm GeV}^2$.

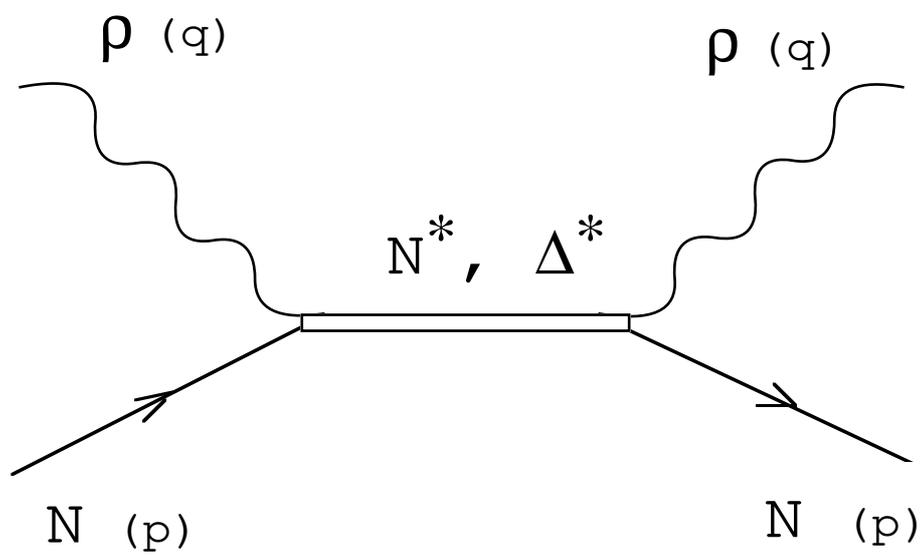
\item{Fig.2:} A schematic illustration of the
 $V-N$ scattering with s-channel nucleon resonances.

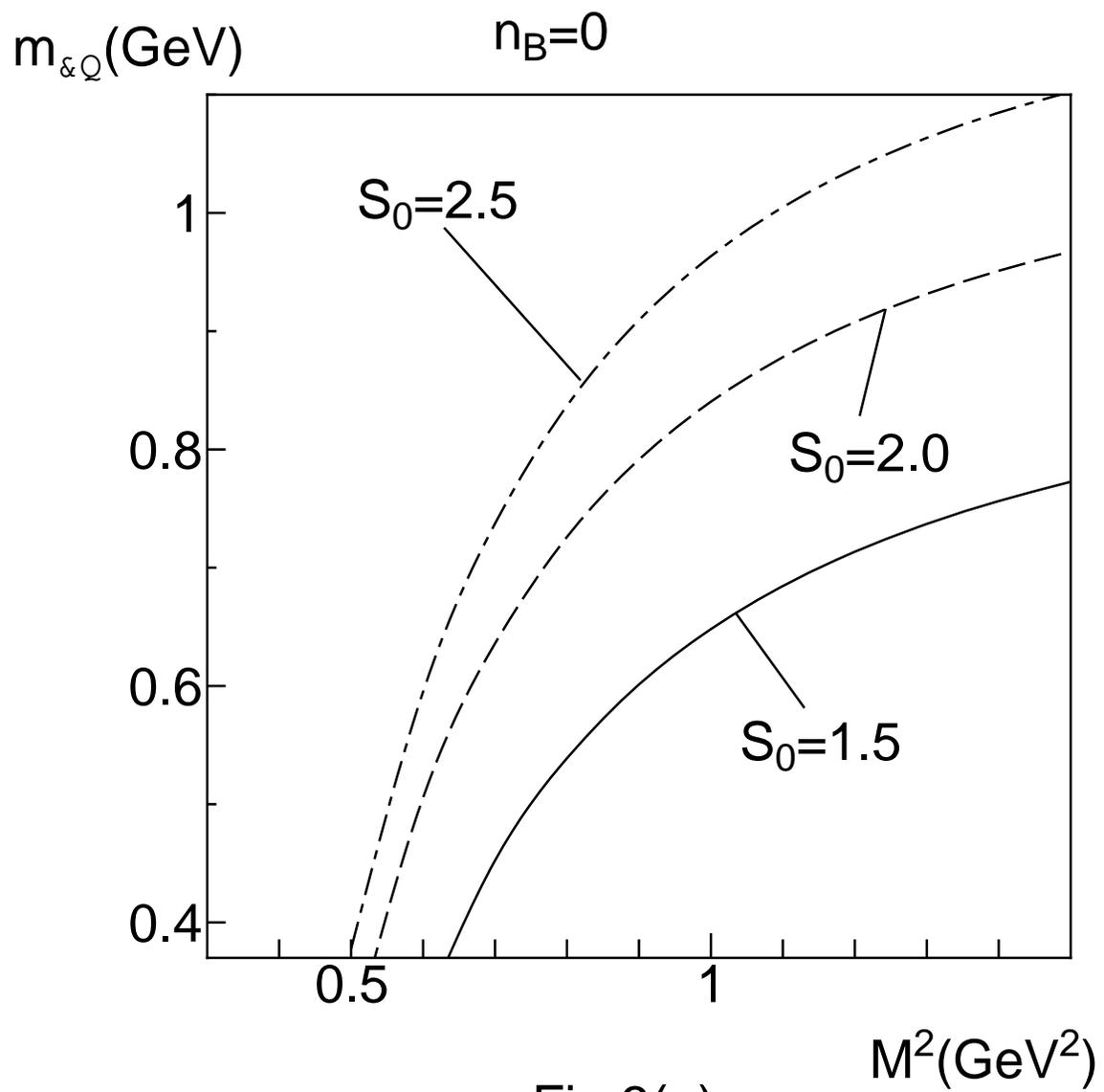
\item{Fig.3:} (a) Borel curves for $m_{_V}(M^2)$ at zero density 
 using $\Pi^R_{\mu \mu}$ with several different values of 
 $S_0$ in GeV$^2$ unit.
 (b) Borel curves for $m_{_V}(M^2)$ at nuclear matter density
  using $\Pi^R_{\mu \mu}$  with several different values of  
$S_0$ in GeV$^2$ unit.

\end{description}


\begin{thebibliography}{99}
\bibitem{tchan} There are also $t-$channel meson exchanges in the
 process. One of them is the scalar-meson  ($\sigma$) exchange
 which gives energy independent contribution to the amplitude and
 is more relevent to the vector-meson mass shift in nuclear matter.
\bibitem{further}
For the readers who has not been convinced yet, we give here another
 argument.  The FESR can be constructed by taking
 certain number of derivative of the real part of
 $\Pi^R_{\mu \nu}(Q^2)$ \cite{PIV}.
 If one start with $\Pi^R(Q^2)
 \sim \ln Q^2 + a_1/Q^2 + a^2/Q^4 + a_3/Q^6 $
  and apply $Q^2 \partial / \partial Q^2$,
 one obtains the canonical FESR (\ref{FESR}). The local duality
 relation in (\ref{FESR}) is nothing but the statement that $a_1 =0$.
 On the other hand,
 If one start with $\Pi^R_{\mu \mu} (Q^2) \sim
 Q^2 \Pi^R(Q^2) \sim Q^2 \ln Q^2 + a_1 + a^2/Q^2 + a_3/Q^3 $,
 any derivative with respect to $Q^2$ will loose the information of
 $a_1$. This is equivalent to the missing of the local duality 
 relation
 or the ambiguity of $s\delta(s)$ in ${\rm Im} \Pi^R_{\mu \mu}(s)$.
 TH thanks Dr. K. Tanaka for  discussion on this point.
\end{thebibliography}
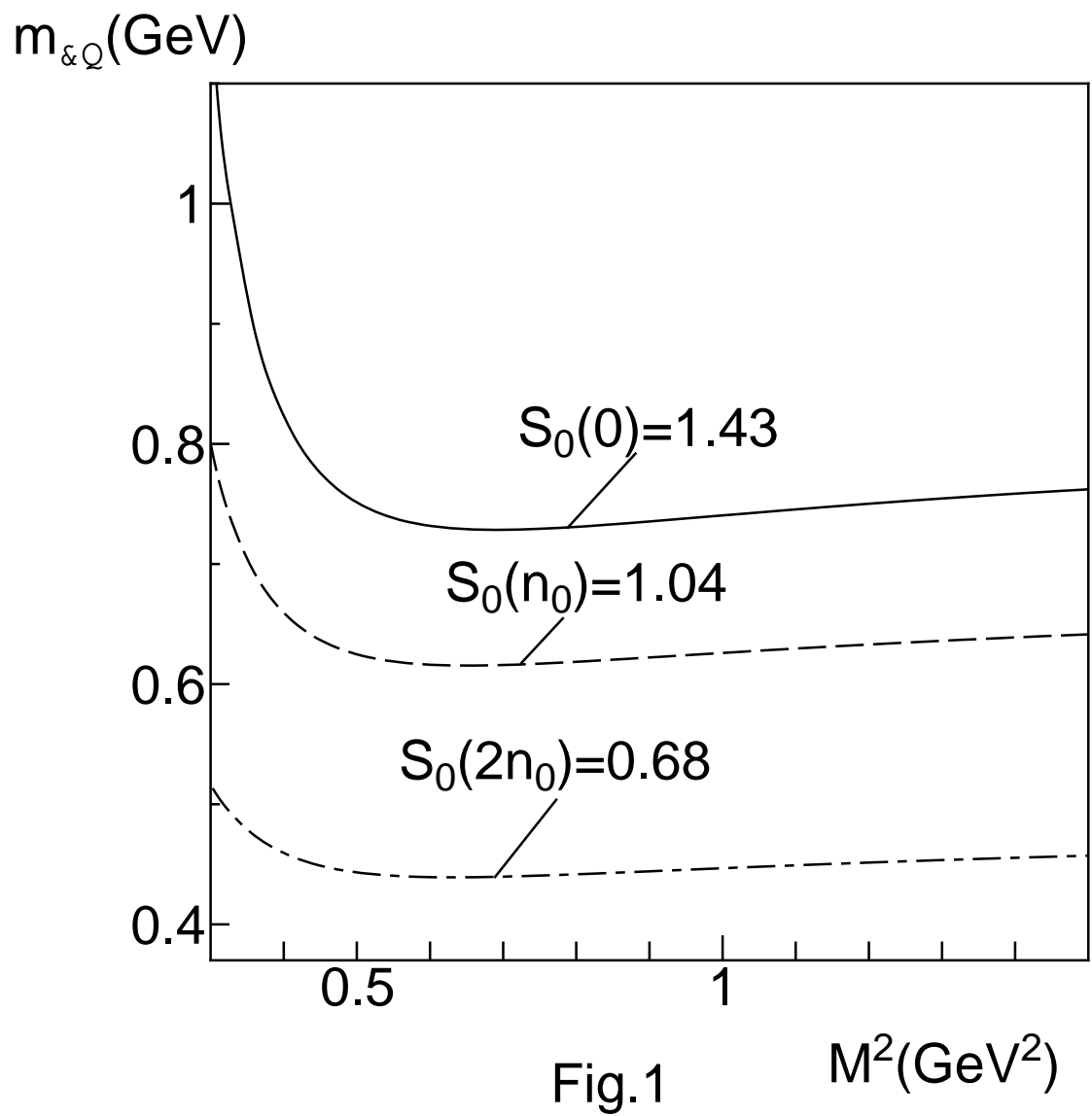
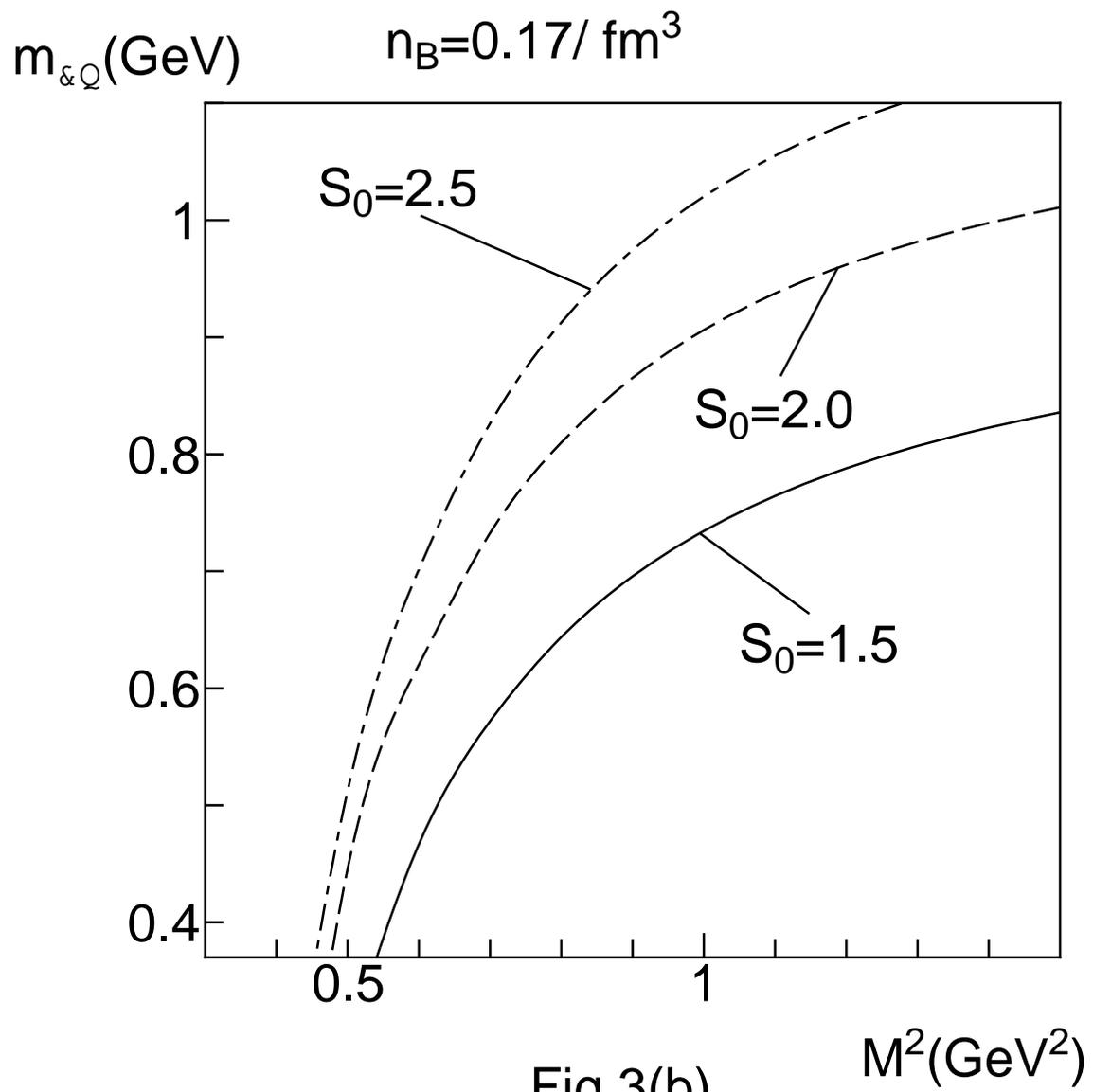
\end{document}